# Where the SRAM Goes: Per-Symbol Attribution of Static Memory in FreeRTOS-C and Embassy-Rust IoT Firmware on a 64 KB RISC-V Microcontroller

**Romi Fadli, Jan Everhard Riwurohi, Wahyu Ramadani Putri Utami, Zulfa Malika Amana**
*Magister Ilmu Komputer, Universitas Budi Luhur, Jakarta, Indonesia*
*Corresponding author: Romi Fadli (2511600872@student.budiluhur.ac.id)*

**Abstract** Selecting a firmware stack for a memory-constrained microcontroller is usually guided by aggregate binary sizes, which report how much SRAM a build consumes but not where it goes or why. We address that gap with a per-symbol attribution method that decomposes static RAM into architectural categories — scheduler heap, network buffers and pools, task stacks, and linker-reserved regions — by bounding sized ELF symbols to the .data, .bss and .stack section ranges. The method is validated by reconciling every category sum exactly against the size-based totals for all sixteen builds. We apply it to eight cumulative firmware milestones, from a bare task scaffold to a full MQTT-over-Ethernet client, implemented twice — in FreeRTOS-C and in Embassy-Rust — on the CH32V203RBT6, a RISC-V microcontroller with 64 KB SRAM and an on-chip Ethernet MAC and PHY, with functional parity verified on hardware at every milestone. The decomposition shows that both stacks spend a nearly identical share of their RAM on networking (63.8% versus 63.9%), yet the absolute cost differs by 4.4 times: 40,719 bytes for lwIP against 9,316 bytes for smoltcp. FreeRTOS task stacks grow linearly at roughly 1 KB per task while Embassy async state grows with state-machine complexity. We also report that lwIP could not be integrated under an all-static discipline on this platform, requiring a 20 KB arena that smoltcp does not. The measurement harness, the attribution tool, and the sixteen firmware implementations are released as artefacts.

## 1 Introduction

### 1.1 Background

The Internet of Things (IoT) industry has driven increasing demand for resource-constrained microcontrollers capable of handling real-time networking workloads. RISC-V, a free and open reduced instruction set architecture, has gained considerable traction in this segment owing to its royalty-free licensing, competitive performance per watt, and a growing ecosystem of open-source toolchains [1], [11]. Among the most capable low-cost RISC-V microcontrollers currently available is the CH32V203RBT6, manufactured by Nanjing Qinheng Microelectronics Co., Ltd. (WCH). This device integrates a 32-bit QingKe V4B core running at up to 144 MHz, 128 KB Flash, 64 KB SRAM, and — uniquely for its price tier — a fully integrated on-chip 10 Mb/s Ethernet MAC and PHY [3]. This last feature eliminates the need for an external PHY component, making it attractive for cost-sensitive IoT applications that require network connectivity, such as industrial sensor nodes, home automation bridges, and MQTT-based data loggers.

Delivering a complete networked application on a device with 64 KB SRAM, however, presents a significant memory-management challenge. The dominant firmware paradigm for such devices combines the C programming language with a real-time operating system (RTOS). FreeRTOS is the most widely deployed open-source RTOS for microcontrollers [2], and its combination with the lwIP lightweight IP stack forms the de facto standard for networked RISC-V and ARM Cortex-M firmware. Yet this combination carries a substantial static memory overhead: the FreeRTOS scheduler, lwIP buffer pools, and C runtime together consume a significant fraction of available SRAM before the application logic itself is accommodated [5]. As IoT firmware grows in complexity — adding DNS resolution, ICMP, NTP synchronization, and MQTT Quality-of-Service — this overhead scales non-linearly, and the risk of running out of SRAM on a 64 KB device becomes real [6].

A concurrent development in the embedded systems community is the emergence of Rust as an alternative to C for firmware programming. Rust offers memory safety without a garbage collector through its ownership and borrow-checking type system, which eliminates entire classes of undefined behavior — null pointer dereferences, use-after-free, and buffer overflows — at compile time rather than at runtime [11], [17]. Embassy is an asynchronous Rust framework specifically designed for bare-metal microcontroller firmware [9]. Embassy eliminates the need for a traditional RTOS scheduler by leveraging Rust's async/await cooperative tasking model, implemented as a zero-cost abstraction at the language level. This design philosophy results in dramatically smaller runtime overhead compared to a preemptive RTOS: no thread stacks, no heap, and no scheduler data structures need to be allocated in SRAM [9].

### 1.2 Problem Statement

Despite the theoretical memory advantages of Embassy-Rust, empirical evidence on resource-constrained RISC-V microcontrollers — particularly those with integrated Ethernet — remains limited. Prior comparative studies have been conducted on ARM Cortex-M platforms [11], [10] or at a conceptual level without systematic milestone-by-milestone measurements across a full IoT networking stack. The question of how much SRAM Embassy-Rust actually saves compared to FreeRTOS-C on a 64 KB device, and whether the savings are sufficient to make otherwise impractical applications feasible, has not been answered for the CH32V203 class of RISC-V microcontrollers.

This paper addresses that gap through a structured empirical investigation. A set of eight firmware milestones was designed, each building incrementally on the previous: from a minimal idle task (M00) through GPIO, USART, ADC, Ethernet with DHCP, DNS, NTP, and finally MQTT with QoS 0 (M07). Both a FreeRTOS-C and an Embassy-Rust implementation were developed for each milestone, compiled with size-optimizing compiler flags, and measured using ELF binary analysis tools that report precise static memory allocation for text, data, and BSS segments. To the best of the authors' knowledge, this is the first study to conduct a quantitative milestone-by-milestone comparison of FreeRTOS-C and Embassy-Rust on a RISC-V microcontroller with an integrated on-chip Ethernet MAC/PHY, and to hold both implementations to a no-dynamic-allocation policy at the application level throughout.

### 1.3 Contributions

This paper makes the following contributions:

1. **A per-symbol static memory attribution method.** We decompose static RAM into architectural categories (scheduler heap, network buffers and pools, task stacks, linker-reserved regions) by bounding sized ELF symbols to the .data, .bss and .stack section ranges. Unlike aggregate size reporting, the method explains where SRAM goes; unlike hand estimation, every category sum reconciles exactly with the size-based total for all sixteen builds.
2. **An empirical characterisation of the networking memory cliff.** Applying the method across eight cumulative milestones shows that both stacks commit a nearly identical proportion of RAM to networking (63.8% versus 63.9%) while differing 4.4-fold in absolute cost, and that FreeRTOS task stacks scale linearly with task count whereas Embassy async state scales with state-machine complexity. This separation predicts how the comparison moves on other platforms.
3. **A reproducible measurement harness and firmware corpus.** Sixteen functionally verified firmware implementations (eight milestones, two stacks) and an automated build-and-measure harness that emits per-milestone binary metrics with both cold- and warm-cache build timings.
4. **A diagnosed toolchain defect and its workaround.** The Embassy HAL for this device does not build out of the box: 32-bit general-purpose timer support is compile-gated to two chip families, yet the CH32V203RBT6 instantiates such a timer, so the generated code references a conditionally absent enum variant. We diagnosed this, patched it locally to obtain a working build, and report it as an instance of the ecosystem-maturity gap that distinguishes the two stacks in practice. An equivalent fix was independently proposed upstream while this work was in preparation.

The remainder of the paper is organised as follows. Section 2 positions the work against prior art. Section 3 describes the hardware platform, the two toolchains, the eight-milestone protocol, and the attribution method. Section 4 reports the measurements and the decomposition. Section 5 concludes.

## 2 Related Work

### 2.1 RISC-V in Embedded and Real-Time Systems

RISC-V has emerged as a credible architecture for embedded real-time applications [1]. Balas et al. [12] introduced CV32RT, an enhancement to the open-source CV32E40P RISC-V core that achieves interrupt latency as low as 6 clock cycles through a custom fast-IRQ extension to the RISC-V Core Local Interrupt Controller (CLIC). Their work demonstrated that RISC-V cores can achieve interrupt performance comparable to ARM Cortex-M and Infineon TriCore, a prerequisite for reliable real-time networking. Behnke et al. [4] addressed the challenge of deterministic IP packet processing on embedded real-time devices, showing that without priority-aware modifications to the network driver the reception of bursty IP traffic degrades real-time task deadlines. Their work is directly relevant to the Ethernet-over-DHCP and MQTT milestones examined in this paper, where lwIP's interrupt handling and FreeRTOS task priorities interact under load. The ESP32-C3, another popular RISC-V IoT microcontroller [14], has been used as a baseline reference for comparing Ethernet-capable embedded Linux versus RTOS deployments, but it lacks the integrated on-chip Ethernet MAC/PHY present on the CH32V203, making it a less direct comparison for the networking milestones studied here.

### 2.2 FreeRTOS and lwIP Architecture

The FreeRTOS kernel provides task scheduling, inter-task communication (queues, semaphores, event groups), and software timers [2]. The kernel was designed to be minimal but configurable: heap management, tick rate, stack sizes, and the presence or absence of optional features are all controlled at compile time through FreeRTOSConfig.h. In a static-allocation-only configuration, FreeRTOS eliminates dynamic heap allocation entirely, committing all task stacks and kernel data structures to the BSS section at link time [2]. This configuration was used throughout the present study. The integration of FreeRTOS with the lwIP network stack requires that lwIP buffer pool sizes be specified statically at compile time; these pool tables dominate SRAM consumption at the networking milestones [6]. Mannella et al. [15] studied FreeRTOS dependability under fault injection, characterizing the vulnerability of scheduler data structures (task control blocks, ready lists) to radiation-induced soft errors, which motivates the interest in smaller, simpler software stacks for safety-critical IoT deployments.

### 2.3 Embedded Rust and Embassy

Rust's memory safety guarantees have attracted growing interest from the embedded systems community as a means of reducing firmware defects without incurring garbage-collection overhead [11], [17]. Vandervelden et al. [17] provided a comprehensive overview of embedded Rust operating systems and frameworks — including RTIC, Tock OS, Hubris, Drone OS, and Embassy — comparing their concurrency models, hardware abstraction layer designs, and ecosystem maturity. Among these, Embassy was identified as the most actively maintained framework offering a complete asynchronous networking stack (embassy-net over smoltcp) that targets both ARM Cortex-M and RISC-V. Frank et al. [18] presented Ariel OS, a multicore-capable embedded Rust OS that incorporates Embassy's asynchronous executor and provides performance benchmarks across ARM, Espressif Xtensa, and RISC-V targets. Seidel and Beier [16] examined the use of Rust in space-qualified embedded systems, concluding that Rust's compile-time safety guarantees are directly applicable to safety-critical microcontroller firmware with no measurable runtime overhead. Molz et al. [20] mapped MISRA-C++ coding guidelines to their Rust-language equivalents (MISRust), providing evidence that the Rust type system enforces a large subset of MISRA-C++ safety rules statically and without annotation overhead.

### 2.4 Comparative Studies of C and Rust Firmware

Sharma et al. [11] performed the first systematic study of the current state of Rust for embedded systems, surveying 225 developers and analysing 6,408 Rust embedded software repositories. They found that while the Rust ecosystem has matured significantly, challenges remain in SAST tooling for embedded targets and interoperability with existing C codebases. Thapa et al. [10] reported an industrial case study in which two teams — one in C and one in Rust over Ariel OS — developed the same IoT sensor logger firmware concurrently on STM32 hardware. Their analysis showed no strong advantage for C over Rust in either memory footprint or execution speed on ARM Cortex-M, and that Ariel OS provided a smaller runtime footprint than the reference baremetal C stack. The present study extends this line of investigation to the RISC-V architecture and to a device with integrated Ethernet MAC/PHY, providing the first milestone-by-milestone comparison that isolates the memory cost of each networking layer. Tock OS [22], a Rust-based embedded OS designed for extreme isolation and correctness, has been deployed on over 10 million production devices, providing evidence of Rust's long-term viability for MCU firmware in commercial products. Yoshimura et al. [7], [8] explored Rust component frameworks (TECS) for RTOS systems, showing that Rust's type system can enforce inter-component contracts that previously required dynamic runtime checking in C-based OS frameworks.

### 2.5 Research Gap

No prior published work has conducted a quantitative eight-milestone comparison of FreeRTOS-C versus Embassy-Rust on a RISC-V microcontroller with an integrated on-chip Ethernet MAC/PHY and only 64 KB of SRAM. The CH32V203RBT6 represents a class of affordable RISC-V IoT devices that are too resource-constrained for lwIP's default configuration yet capable enough to host a full MQTT stack under the right firmware framework. This paper fills that specific empirical gap.

## 3 Materials and Methods

### 3.1 Hardware Platform

All experiments were conducted on the CH32V203RBT6-R1 development board, which hosts the CH32V203RBT6 microcontroller manufactured by Nanjing Qinheng Microelectronics Co., Ltd. The core specifications relevant to this study are listed in Table 1.

**Table 1** CH32V203RBT6 Hardware Specifications

| Parameter | Value |
|---|---|
| CPU core | QingKe V4B (RISC-V RV32IMAC) |
| Maximum clock | 144 MHz (operated at 120 MHz in this study) |
| Flash memory | 128 KB (131,072 bytes) at 0x00000000 |
| SRAM | 64 KB (65,536 bytes) at 0x20000000 |
| Ethernet | Integrated 10BASE-T MAC and PHY (no external PHY) |
| High-speed crystal (HSE) | 32 MHz external (board-mounted) → PLL → 120 MHz system clock |
| Low-speed crystal (LSE) | 32.768 kHz external (board-mounted, RTC) |
| GPIO used | PA0 (LED1), PA1 (LED2/PWM TIM2-CH2), PB0 (ADC IN8), PB2 (button EXTI2) |
| Debug interface | WCH-Link (RISC-V OpenOCD, CMSIS-DAP) on-board |
| Package | LQFP64 |

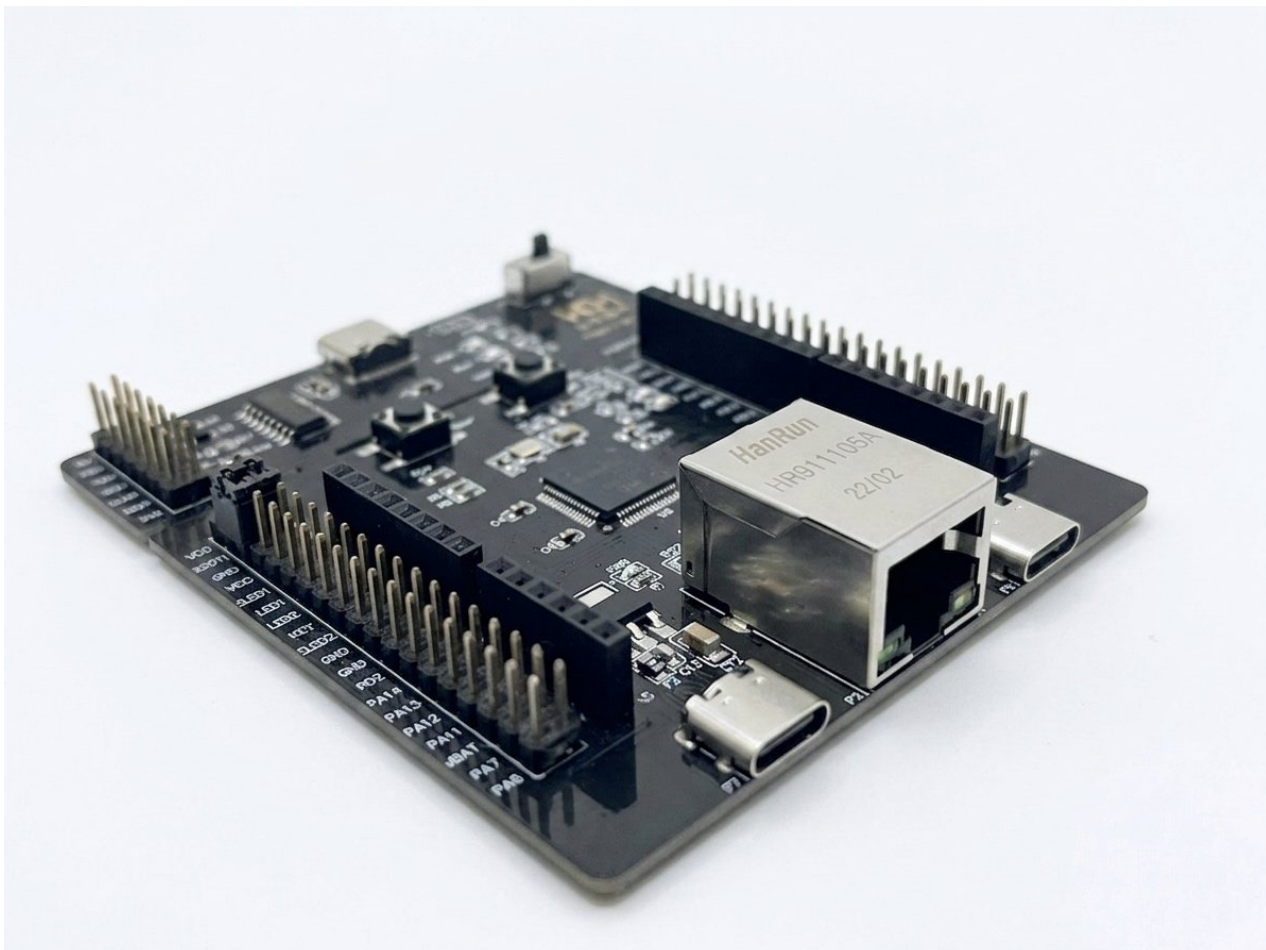

**Fig. 1** CH32V203RBT6-R1 development board used in this study: QingKe V4B RISC-V core, 64 KB SRAM, 128 KB Flash, integrated 10BASE-T Ethernet MAC/PHY, WCH-Link on-board debugger.

The 120 MHz system clock was derived by configuring the PLL to multiply the 32 MHz HSE crystal by a factor of 3.75. The integrated Ethernet MAC/PHY operates in 10BASE-T half-duplex mode and requires no external transformer for short-distance bench connections.

### 3.2 FreeRTOS-C Toolchain Configuration

The FreeRTOS-C firmware was developed using MounRiver Studio 2 (v2.x) with the GNU Make build system. Table 2 details the toolchain configuration.

**Table 2** FreeRTOS-C Toolchain

| Component | Version / Setting |
|---|---|
| Compiler | riscv-none-embed-gcc 8.2.0 |
| Architecture flags | -march=rv32imac -mabi=ilp32 |
| Optimization | -Os (optimize for code size) |
| FreeRTOS version | V10.4.6 |
| lwIP version | 2.1.3 |
| Startup | startup_ch32v20x_D8.S |
| Linker script | Link.ld (128 KB Flash, 64 KB RAM, stack 4,096 bytes) |
| Heap policy | Static for all application objects; bounded arena for lwIP port |
| Tick rate | 500 Hz (configTICK_RATE_HZ=500) |
| Min stack size | 128 words = 512 bytes (configMINIMAL_STACK_SIZE=128) |
| Max priorities | 7 (configMAX_PRIORITIES=7) |
| Total heap | 0 B (M00-M04); 14 KB (M05); 20 KB (M06-M07), lwIP port only |

### 3.3 Embassy-Rust Toolchain Configuration

The Embassy-Rust firmware was compiled using the stable Rust toolchain for the riscv32imac-unknown-none-elf target. Table 3 details the Rust configuration.

**Table 3** Embassy-Rust Toolchain

| Component | Version / Setting |
|---|---|
| Rust toolchain | stable (edition 2021) |
| Compilation target | riscv32imac-unknown-none-elf |
| Optimization | opt-level = "z" (minimize binary size) |
| Link-time optimization | lto = true |
| Codegen units | codegen-units = 1 |
| Embassy executor | embassy-executor 0.9.1 |
| Embassy networking | embassy-net 0.7 (smoltcp 0.12) |
| Embassy time | embassy-time 0.5.0 |
| Embassy sync | embassy-sync 0.7 |
| Static data structures | heapless 0.8.0, static_cell 2 |
| HAL | ch32-hal (patched local fork for CH32V203 support) |
| RISC-V support | qingke 0.7 |
| Global allocator | None (no heap — no GlobalAlloc implementation) |

Both implementations were held to a no-dynamic-allocation policy for all application-level objects, in line with NASA JPL Rule 3 [24] and MISRA-C Rule 21.3 [25], which prohibit dynamic allocation after initialisation to prevent fragmentation and to permit static worst-case analysis. Every task, queue, semaphore and timer in the FreeRTOS-C application code is created through the xXxxCreateStatic() family, and the Embassy-Rust workspace links no GlobalAlloc implementation at all, so an attempted heap allocation there is a compile-time error.

The two stacks could not, however, be held to the same policy at the network layer, and this asymmetry is itself a result rather than a caveat. The lwIP port layer allocates its mailboxes, mutexes and its internal thread through the FreeRTOS dynamic API (sys_mbox_new, sys_mutex_new, sys_thread_new), which the static API cannot satisfy. Integrating lwIP therefore required enabling configSUPPORT_DYNAMIC_ALLOCATION and reserving a bounded arena from the milestone at which networking is introduced: 14 KB at M05, raised to 20 KB at M06-M07. Milestones M00-M04 remain strictly heap-free (configTOTAL_HEAP_SIZE = 0). The arena is reserved exclusively for the lwIP port; no application code calls pvPortMalloc after initialisation. Embassy-Rust required no equivalent concession: smoltcp accepts caller-provided static buffers, so the Embassy builds remain allocator-free at every milestone. Because the arena is a fixed reservation in BSS it remains fully visible to static analysis, and it is reported as its own category in Section 4.

### 3.4 Eight-Milestone Methodology

A structured milestone progression was defined to isolate the incremental memory cost of each firmware subsystem. Table 4 describes each milestone.

**Table 4** Eight-Milestone Firmware Progression

| ID | Label | Cumulative Subsystems Added |
|---|---|---|
| M00 | Baseline | Minimal scheduler, one idle task, no peripherals |
| M01 | GPIO + LED + Button | GPIO output (LED1), EXTI2 interrupt (button), digital input |
| M02 | + USART | USART1 at 115200 baud, printf-style logging |
| M03 | + ADC + PWM | ADC1 channel 8 (PB0), TIM2-CH2 PWM on PA1 |
| M04 | + Internal Temp Sensor | ADC1 internal temperature channel |
| M05 | + Ethernet + DHCP | Ethernet MAC/PHY init, lwIP/smoltcp, DHCP client |
| M06 | + DNS + ICMP + NTP | DNS resolution, ICMP ping, NTP time sync |
| M07 | + MQTT QoS 0 | MQTT client, subscribe and publish, QoS 0 |

Each milestone was compiled as a distinct build artifact. No milestone modified the configuration of a previously validated milestone; only additive code was introduced. For FreeRTOS-C, each milestone consisted of a separate main.c with the appropriate peripheral drivers and lwIP configuration. For Embassy-Rust, each milestone corresponded to a separate binary target in the Cargo workspace.

### 3.5 Memory Measurement Protocol

Static memory consumption was measured using the riscv-none-embed-size utility (GNU Binutils) applied to the compiled ELF output of each milestone build. The size command reports the sizes of three ELF segments: text (executable code and read-only constants, contributing to Flash); data (initialized global/static variables, occupying both Flash and RAM at runtime); and bss (zero-initialized global/static variables, occupying RAM only). Flash consumption was computed as: text + data. RAM consumption was computed as: data + bss. All measurements were performed on release-mode (optimized) builds. Build time is the wall-clock time of the build command alone (make, or cargo build --release), excluding the preceding clean step, and was measured in two regimes. In the cold regime every dependency is rebuilt from scratch (make clean for FreeRTOS-C; cargo clean of all crates for Embassy-Rust). In the warm regime the Embassy dependency cache is kept and only the firmware crate is rebuilt (cargo clean

-p firmware); FreeRTOS-C has no separate dependency cache, so its warm build is again make clean followed by make. Both tools were invoked with their default parallelism: make without -j (one compilation unit at a time) and cargo with one job per hardware thread. Each regime was run five times, sequentially and never overlapping, on the same machine (Lenovo Legion 5 15ARH7 laptop with an AMD Ryzen 7 6800H, 8 cores/16 threads, 16 GB DDR5 RAM, and NVMe Gen 4 storage), and the median of the five runs is reported.

### 3.6 Measurement Validity

To ensure reproducibility: each build was executed from the clean state defined for its regime in Section 3.5 before measurement; binary sizes were checked to be byte-identical across all repeated runs; compiler flags were verified to be identical between measurement runs; the same ELF analysis script was applied to both toolchain outputs; all FreeRTOSConfig.h parameters were fixed across milestones M00-M07, with only the set of compiled source files changing at each milestone; and the Embassy Cargo workspace pinned all dependency versions in Cargo.lock.

### 3.7 Functional Verification

To confirm that both implementations provide equivalent functionality at every milestone, each was verified on the same CH32V203RBT6-R1 board against the criteria listed in Table 5. Verification was performed manually for each framework before recording the memory measurements, ensuring that only functionally correct builds are included in the results.

**Table 5** Functional Verification Criteria per Milestone

| Milestone | Verification Method | Expected Outcome |
|---|---|---|
| M01 | Visual inspection | LED1 toggles on each button press; LED2 on at startup |
| M02 | Serial terminal (115,200 baud) | Task info printed every 1 s; uptime counter increments |
| M03 | Multimeter / oscilloscope | PWM duty cycle changes proportionally to sensor input |
| M04 | Serial terminal | Internal temperature reading 25–40 °C; updates every 1 s |
| M05 | Router DHCP lease table | Device acquires an IP address within 5 s of power-on |
| M06 | ping from PC; serial terminal | ICMP echo reply < 100 ms; NTP time visible on serial |
| M07 | MQTT broker log (Mosquitto) | Device connects, subscribes, and publishes QoS 0 messages |

Both implementations passed all seven verification criteria. Functional parity is therefore confirmed across the full M00–M07 progression.

### 3.8 Per-Symbol Static Memory Attribution

Segment totals report how much SRAM a build consumes but not what consumes it. To recover that, a second measurement pass attributes each byte of static RAM to an architectural category. The procedure is: (i) read the .data, .bss and .stack section headers from the linked image, which are authoritative for what occupies SRAM; (ii) enumerate sized symbols and retain only those whose address falls inside those ranges; (iii) assign each retained symbol to a category by name pattern; and (iv) report any SRAM reserved by the linker script but carrying no sized symbol as its own category rather than discarding it.

Step (ii) is what makes the result trustworthy. Filtering on symbol type alone over-counts, because a symbol may carry a BSS or data type code while residing outside those sections; bounding by address range removes that error. Step (iv) captures the 4 KB main and interrupt stack reserved through __stack_size, which occupies real SRAM that the application cannot use but which no symbol describes.

The method is self-validating: because the categories partition the same address ranges the size utility measures, their sum must equal the segment total. This identity holds exactly for all sixteen builds, which distinguishes the resulting figures from the component estimates commonly reported in comparisons of this kind. Categories are architectural rather than stack-specific, so the same taxonomy applies to both implementations and the comparison remains like-for-like. The implementation is released with the firmware corpus.

## 4 Results and Discussion

### 4.1 Overview of Build Metrics

Table 6 summarises the static memory measurements for both firmware stacks across all eight milestones. Device RAM and Flash capacities are 65,536 bytes (64 KB) and 131,072 bytes (128 KB), respectively. Figures 2-5 visualise the data as line and bar charts for easier interpretation of growth trends.

**Table 6** Static Memory Measurements: FreeRTOS-C vs. Embassy-Rust

| M | Description | RTOS RAM | RTOS % | Emb RAM | Emb % | RTOS Flash | RTOS % | Emb Flash | Emb % |
|---|---|---|---|---|---|---|---|---|---|
| M00 | Baseline | 6,928 | 10.6% | 556 | 0.8% | 9,372 | 7.2% | 4,098 | 3.1% |
| M01 | +GPIO/LED/btn | 8,132 | 12.4% | 612 | 0.9% | 14,684 | 11.2% | 5,292 | 4.0% |
| M02 | +USART | 9,240 | 14.1% | 1,232 | 1.9% | 15,104 | 11.5% | 11,624 | 8.9% |
| M03 | +ADC+PWM | 11,464 | 17.5% | 1,360 | 2.1% | 16,800 | 12.8% | 13,170 | 10.0% |
| M04 | +Int. temp | 12,652 | 19.3% | 1,488 | 2.3% | 18,140 | 13.8% | 14,740 | 11.2% |
| M05 | +Ethernet+DHCP | 51,272 | 78.2% | 9,424 | 14.4% | 61,824 | 47.2% | 40,844 | 31.2% |
| M06 | +DNS+ICMP+NTP | 61,208 | 93.4% | 11,752 | 17.9% | 67,224 | 51.3% | 56,400 | 43.0% |
| **M07** | **+MQTT QoS 0** | **63,776** | **97.3%** | **14,568** | **22.2%** | **76,824** | **58.6%** | **73,164** | **55.8%** |

### 4.2 RAM Consumption Analysis

Figure 2 shows the cumulative RAM trajectory from the baseline through the full MQTT stack. FreeRTOS-C (solid red) approaches the 64 KB capacity ceiling by M07, while Embassy-Rust (dashed blue) remains well below 25 KB.

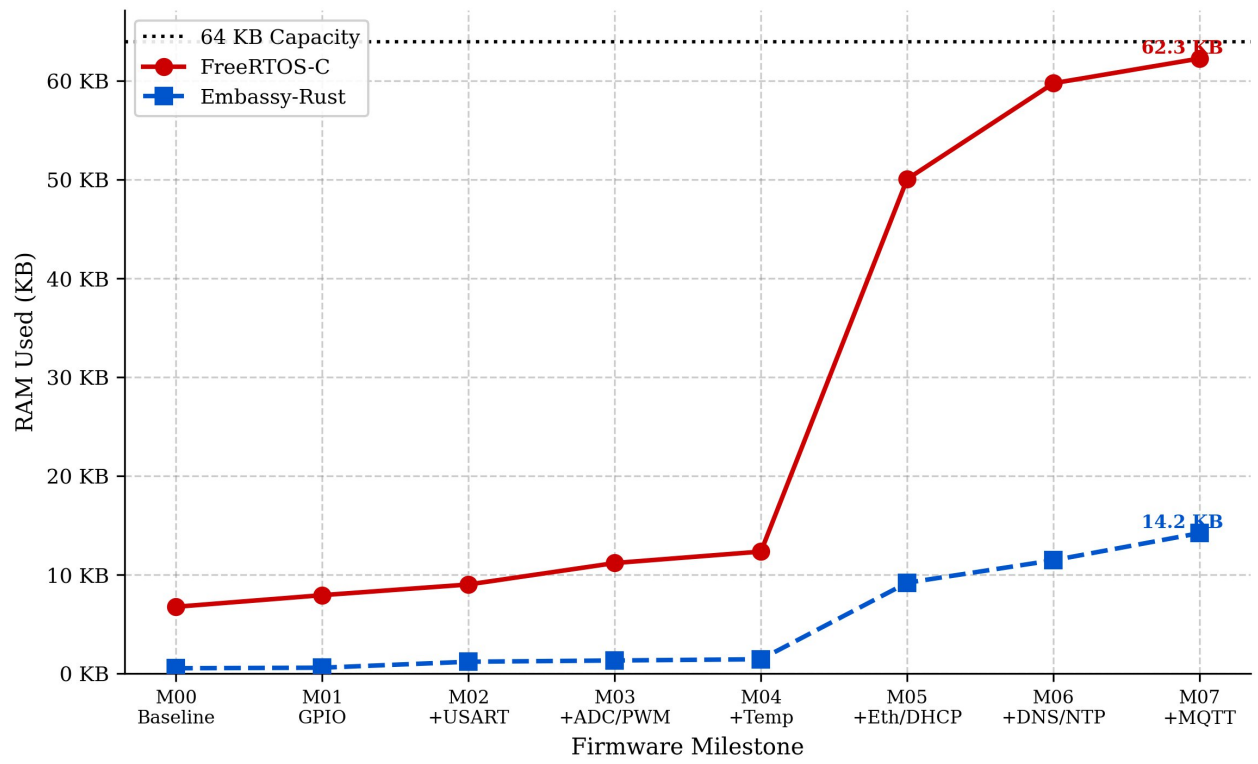

**Fig. 2** Cumulative RAM consumption across milestones M00-M07: FreeRTOS-C (solid red circles) vs. Embassy-Rust (dashed blue squares) vs. 64 KB device capacity (dotted black).

Figure 3 shows the per-milestone incremental RAM growth. The dominant transition is M04->M05 where FreeRTOS-C grew by 38,620 bytes (lwIP pool table) while Embassy-Rust grew by only 7,936 bytes.

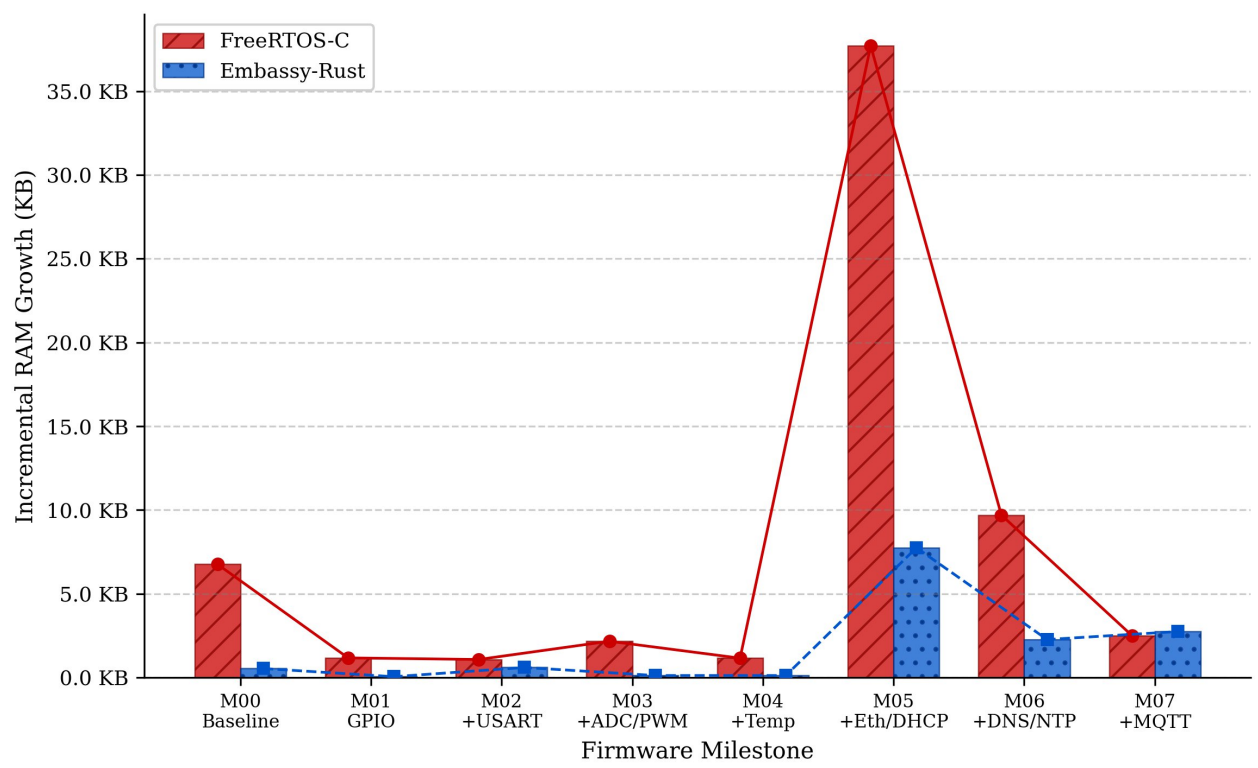

**Fig. 3** Incremental RAM growth per milestone: FreeRTOS-C (red bars, hatched) vs. Embassy-Rust (blue bars, dotted pattern) with trend lines overlaid. The M05 bar represents the Ethernet+DHCP subsystem introduction.

#### *4.2.1 Pre-Ethernet Milestones (M00-M04)*

At the baseline milestone M00, FreeRTOS-C required 6,928 bytes of RAM compared to Embassy's 556 bytes (12.5x ratio). This gap reflects the FreeRTOS kernel allocating idle task control blocks, ready-list arrays for all priority levels, and tick counter in BSS at startup, while Embassy's executor is a minimal static array of waker slots. The four milestones M01-M04 added GPIO, USART, ADC, and PWM peripherals. FreeRTOS-C RAM grew from 6,928 to 12,652 bytes (+5,724 bytes), while Embassy grew from 556 to 1,488 bytes (+932 bytes). Each FreeRTOS peripheral required at least one task stack (min 512 bytes) and a queue handle; Embassy async futures carry no stack overhead beyond the precisely-sized coroutine state.

### *4.2.2 Ethernet Introduction (M04 to M05)*

The most dramatic transition occurred at M05. FreeRTOS-C RAM jumped from 12,652 to 51,272 bytes (+38,620 bytes, +305%), consuming 59% of total SRAM in a single step. The root cause is lwIP's pool table architecture: the stack pre-allocates all network buffer pools, ARP table entries, UDP/TCP PCBs, and queues as static arrays at startup regardless of active connections [6]. Embassy-Rust's smoltcp stack at M05 grew from 1,488 to 9,424 bytes (+7,936 bytes), because smoltcp uses const-generic static buffers sized exactly to the application declaration.

The composition of the M04-to-M05 delta is reported in Section 4.3, where it is derived from the linked images rather than estimated. In summary, the step introduces 34,351 B of network-related storage in FreeRTOS-C (a 14,336 B lwIP arena plus 20,015 B of buffers and pool tables) against 7,244 B of socket and DMA buffers in Embassy-Rust, together with a further 3,072 B of additional task stacks on the FreeRTOS-C side for the lwIP input task. The remaining difference is kernel-object growth and miscellaneous static data.

### *4.2.3 Full Stack (M05-M07)*

From M05 to M07, adding DNS/ICMP/NTP and MQTT QoS 0 added 12,504 bytes to FreeRTOS-C (total: 63,776 bytes, 97.3%) and 5,144 bytes to Embassy-Rust (total: 14,568 bytes, 22.2%). FreeRTOS-C had only 1,760 bytes free; Embassy-Rust retained 50,968 bytes free (29x more). The absolute RAM reduction at M07 is 49,208 bytes (77.2% reduction).

## 4.3 Where the SRAM Goes: Per-Symbol Attribution

The aggregate totals establish that Embassy-Rust uses less RAM but not why. Applying the attribution method of Section 3.8 to the sixteen linked images resolves each stack's SRAM into architectural categories; Table 7 reports the result. Every column sums exactly to the corresponding size-based total, so the decomposition is a measurement of the shipped binaries, not an estimate.

**Table 7** Per-Symbol Static RAM Attribution by Milestone (bytes). Upper block: FreeRTOS-C. Lower block: Embassy-Rust.

| Category | M00 | M01 | M02 | M03 | M04 | M05 | M06 | M07 |
|---|---|---|---|---|---|---|---|---|
| lwIP port heap | 0 | 0 | 0 | 0 | 0 | 14,336 | 20,480 | **20,480** |
| Network buffers/pools | 0 | 0 | 0 | 0 | 0 | 20,015 | 20,047 | **20,239** |
| Task stacks | 2,048 | 3,072 | 4,096 | 6,144 | 7,168 | 10,240 | 13,312 | **15,360** |
| RTOS kernel objects | 636 | 720 | 804 | 972 | 1,056 | 1,328 | 1,580 | **1,844** |
| Linker-reserved stack | 4,100 | 4,097 | 4,097 | 4,100 | 4,100 | 4,158 | 4,169 | **4,172** |
| Other static data | 144 | 243 | 243 | 248 | 328 | 1,195 | 1,620 | **1,681** |
| **Total** | **6,928** | **8,132** | **9,240** | **11,464** | **12,652** | **51,272** | **61,208** | **63,776** |

| Category | M00 | M01 | M02 | M03 | M04 | M05 | M06 | M07 |
|---|---|---|---|---|---|---|---|---|
| Network buffers/pools | 0 | 0 | 0 | 0 | 0 | 7,244 | 8,508 | **9,316** |
| Async task state | 65 | 121 | 737 | 865 | 961 | 1,593 | 2,625 | **4,585** |
| Other static data | 652 | 708 | 1,336 | 1,661 | 1,813 | 2,525 | 3,589 | **3,750** |
| **Total** | **556** | **612** | **1,232** | **1,360** | **1,488** | **9,424** | **11,752** | **14,568** |

*Note.* Category sums reconcile exactly with the size-based ram_bytes for all sixteen builds. Linker-reserved stack is the 4 KB main/IRQ stack reserved by the linker script, which carries no sized symbol but occupies real SRAM.

Three findings follow, none of them visible in the aggregate figures. First, the networking share is almost identical in relative terms and vastly different in absolute terms. At M07 the network subsystem accounts for 40,719 B in FreeRTOS-C (the 20,480 B lwIP arena plus 20,239 B of buffers and pools) against 9,316 B for Embassy-Rust: 63.8% versus 63.9% of each stack's own RAM, but a 4.4-fold difference in bytes. Both architectures devote nearly the same proportion of their budget to networking; they differ in what that proportion costs.

Second, the two stacks scale along different variables. FreeRTOS-C task stacks grow linearly with task count, from 2,048 B at M00 to 15,360 B at M07 at roughly 1 KB per task, because every task receives a conservatively sized stack whether or not it uses it. Embassy-Rust async task state grows from 65 B to 4,585 B over the same range, scaling with the complexity of each task's generated state machine rather than with task count, since an async task stores only the locals live across await points. This is the structural reason the gap widens as firmware grows, and it is the component of the comparison most likely to transfer to other platforms: it depends on the concurrency model, not on this device.

Third, the lwIP arena is the single largest category in the FreeRTOS-C build at M07 and has no counterpart on the Embassy-Rust side. Its 20,480 B is 31.3% of the device's entire SRAM, committed before any application data. Together with the pool tables this confirms that the cost is architectural: lwIP pre-commits capacity for peak load, whereas smoltcp sizes buffers to the sockets the application declares.

### 4.4 Flash Consumption Analysis

Figure 4 shows cumulative Flash usage and Figure 5 shows per-milestone incremental Flash growth. At pre-Ethernet milestones M00-M04, Embassy-Rust used less Flash due to aggressive link-time dead-code elimination. At M07 the stacks converge: FreeRTOS-C 76,824 bytes (58.6%) vs. Embassy-Rust 73,164 bytes (55.8%), a difference of only 3,660 bytes (4.8%). This convergence reflects that protocol code dominates Flash at high milestones. Flash is therefore not the binding constraint on this device class - SRAM is.

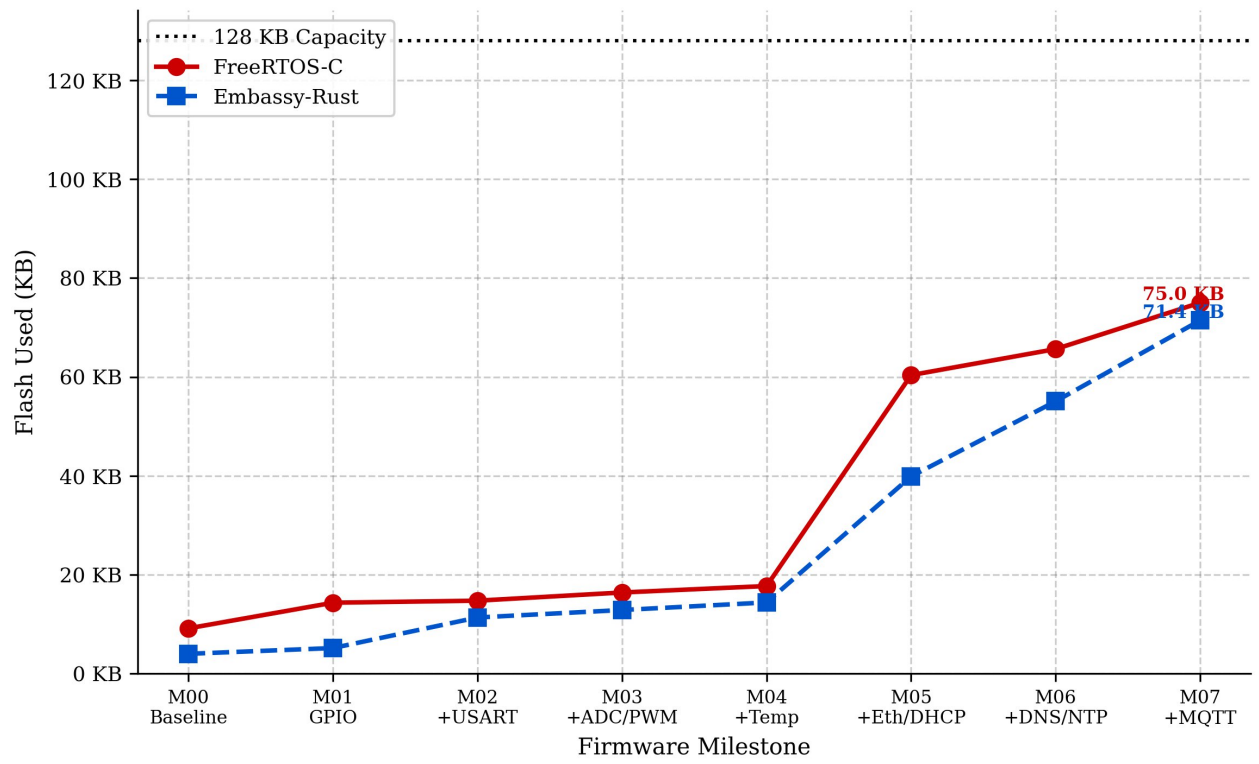

**Fig. 4** Cumulative Flash consumption across milestones M00-M07: FreeRTOS-C (solid red circles) vs. Embassy-Rust (dashed blue squares) vs. 128 KB device capacity (dotted black).

Figure 5 shows that Embassy-Rust's incremental Flash growth at M02 (+USART) is large due to first inclusion of the UART crate, but subsequent milestones add smaller increments. FreeRTOS-C Flash growth is more uniform because lwIP modules are compiled as individual C translation units with minimal cross-module inlining.

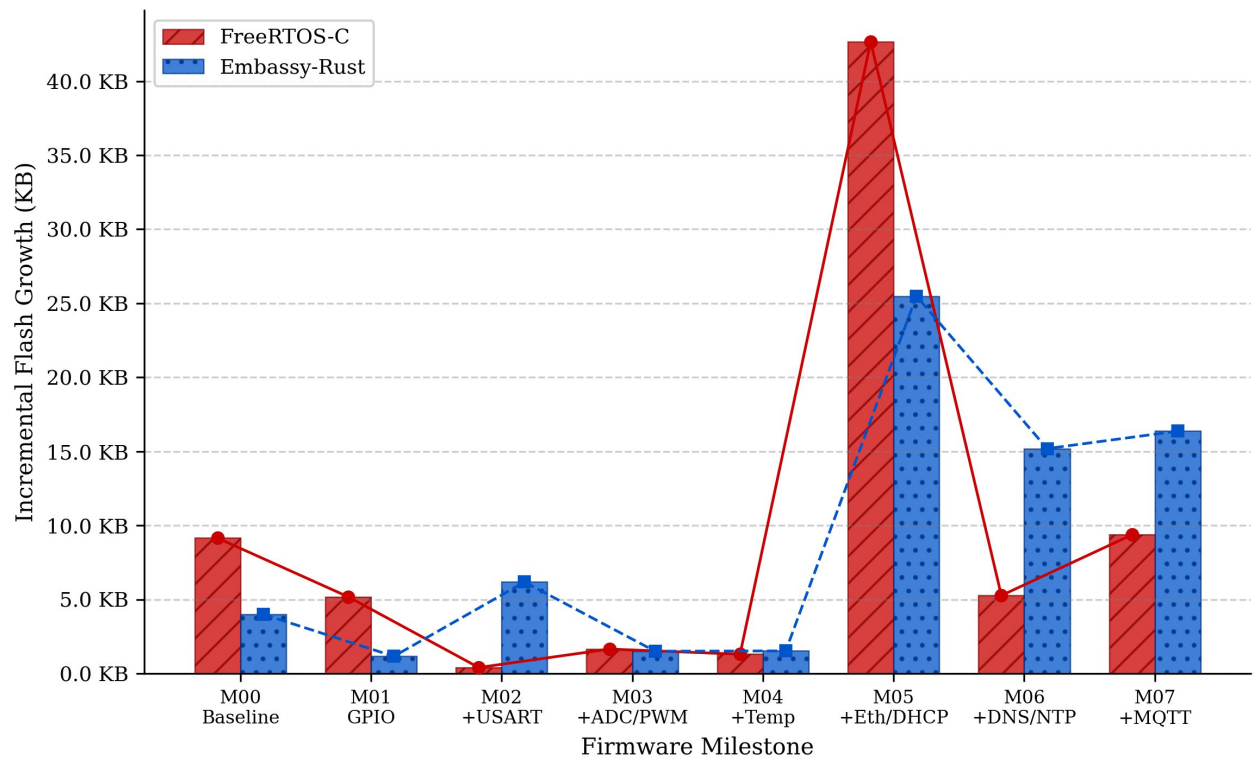

**Fig. 5** Incremental Flash growth per milestone: FreeRTOS-C (red bars, hatched) vs. Embassy-Rust (blue bars, dotted pattern) with trend lines overlaid.

### 4.5 Build Time Comparison

Build time depends strongly on which regime is measured, so both are reported. A cold build compiles every dependency from scratch (make clean; cargo clean for all crates). A warm build recompiles only the firmware crate against an already populated dependency cache (make clean; cargo clean -p firmware), which is the regime a developer experiences during iterative work. Table 8 gives the median of five sequential runs in each regime; the ratio is FreeRTOS-C time divided by Embassy-Rust time, so values below 1 favour FreeRTOS-C.

**Table 8** Build Time by Regime: Cold (Full Dependency Rebuild) and Warm (Incremental), in Seconds

| MS | Cold: FreeRTOS-C | Cold: Embassy-Rust | Cold ratio | Warm: FreeRTOS-C | Warm: Embassy-Rust | Warm ratio |
|---|---|---|---|---|---|---|
| M00 | 2.82 | 7.18 | 0.39x | 2.94 | 0.108 | 27.2x |
| M01 | 2.82 | 7.02 | 0.40x | 2.90 | 0.102 | 28.4x |
| M02 | 2.83 | 6.93 | 0.41x | 2.89 | 0.102 | 28.3x |
| M03 | 2.83 | 7.03 | 0.40x | 2.95 | 0.100 | 29.5x |
| M04 | 2.85 | 7.12 | 0.40x | 2.88 | 0.100 | 28.8x |
| M05 | 7.10 | 7.93 | 0.90x | 7.01 | 0.108 | 64.9x |
| M06 | 7.35 | 8.19 | 0.90x | 7.29 | 0.102 | 71.5x |
| M07 | **7.72** | **8.49** | **0.91x** | **7.60** | **0.101** | **75.3x** |

The two regimes give opposite answers, and conflating them would misrepresent the comparison. Cold, FreeRTOS-C builds faster at every milestone. Embassy-Rust takes 2.4-2.6x longer at the pre-network milestones (M00-M04), where compiling the full Embassy dependency graph dominates a workload that is small for GCC, and 10-12% longer at the networked milestones (0.90-0.91x at M05-M07), where adding the lwIP sources more than doubles the C build time. Warm, Embassy-Rust is 27-75x faster, because Cargo rebuilds only the changed firmware crate against its cached dependencies, whereas the FreeRTOS-C build has no dependency cache and recompiles every translation unit one at a time. The practical reading is therefore about iteration rather than the first build: a cold Embassy-Rust build costs somewhat more than its C counterpart, but each subsequent edit rebuilds in about a tenth of a second. Run-to-run variation is small relative to these differences; the largest single deviation was an Embassy-Rust cold build 21% above its median (M07).

### 4.6 Practical Implications

The results confirm that deploying a complete MQTT-over-Ethernet stack with FreeRTOS-C on the CH32V203RBT6 leaves only 1,760 bytes free. Table 9 lists the approximate minimum RAM requirements of common TLS client implementations and their feasibility given the available headroom for each framework at M07.

**Table 9** TLS Client RAM Requirements vs. Available SRAM Headroom at M07

| TLS Implementation | Min. RAM (approx.) | FreeRTOS-C (1,760 B free) | Embassy-Rust (50,968 B free) |
|---|---|---|---|
| mbedTLS (default config) | 40,000–80,000 B | ✗ Not feasible | ✗ Borderline (tight) |
| mbedTLS (minimal IoT config) | 20,000–30,000 B | ✗ Not feasible | ✓ Feasible (with tuning) |
| WolfSSL (static memory mode) | 10,000–18,000 B | ✗ Not feasible | ✓ Feasible |
| embedded-tls (Rust, no-std) | 15,000–25,000 B | ✗ Not feasible | ✓ Feasible (with tuning) |

As shown in Table 9, no TLS implementation is feasible under FreeRTOS-C on this device: even the most compact option (WolfSSL in static memory mode, ≈ 10,000–18,000 B) requires more than 5× the available 1,760 bytes. With Embassy-Rust, the 50,968 bytes remaining at M07 are sufficient to accommodate WolfSSL or embedded-tls, along with additional sensor drivers and an OTA update buffer. Separately from memory, the warm-build advantage of Embassy-Rust (27-75x across M00-M07, Table 8) shortens the edit-build cycle during firmware development; it does not carry over to first, cold builds, where FreeRTOS-C is faster.

### 4.7 Comparison with Published Work

Table 10 places the results of this study alongside the closest comparable published work: Thapa et al. [10], who conducted a concurrent FreeRTOS-C versus Rust/Ariel OS comparison on the STM32U585 (786 KB SRAM).

**Table 10** RAM Savings: This Study vs. Thapa et al. [10]

| Study | Platform | SRAM (KB) | C/FreeRTOS RAM | Rust/Embassy RAM | Bytes Saved | Reduction |
|---|---|---|---|---|---|---|
| Thapa et al. [10] | STM32U585 | 786 | 44,656 B | 24,640 B | 20,016 B | 44.8% |
| This study | CH32V203RBT6 | 64 | 63,776 B | 14,568 B | 49,208 B | 77.2% |

The RAM saving in this study (77.2%) substantially exceeds that reported by Thapa et al. (44.8%). The primary explanation is platform constraint: on the 786 KB STM32U585, lwIP's pre-allocated pools represent a small fraction of total SRAM, so their absolute cost is less impactful. On the 64 KB CH32V203RBT6, the same pools consume 59% of total SRAM in a single step (M05). Furthermore, the networking scope in this study is broader — full wired Ethernet with DHCP, DNS, ICMP, NTP, and MQTT — versus the SPI-connected MQTT client in Thapa et al., making the lwIP pool cost more prominent.

### 4.8 Functional Parity and Qualitative Observations

Both implementations passed all functional verification tests described in Section 3.7. Identical observable behaviour was confirmed on the same CH32V203RBT6-R1 board at every milestone: LEDs toggled on button press, USART output matched the expected format, ADC readings responded to the sensor input, DHCP leases were obtained within 5 seconds, ICMP ping responded under 100 ms, NTP synchronization was observed, and MQTT messages were published and received by the same broker. Embassy's async cooperative scheduling produces equivalent functional outcomes to FreeRTOS's preemptive multi-tasking for the IoT workloads studied.

Several qualitative differences were noted during development. First, Embassy's Rust type system detected a compile-time error when two async tasks attempted unsynchronised access to the ADC peripheral — a condition that constitutes a silent runtime race in FreeRTOS-C and is not detected by the compiler. Second, FreeRTOS-C benefited from an official WCH-provided kernel port and a well-tested lwIP integration; Embassy required a locally patched ch32-hal fork, reflecting the relative ecosystem maturity for this specific MCU at the time of study. Third, FreeRTOS-C provides runtime introspection APIs (vTaskList(), uxTaskGetStackHighWaterMark()) useful during debugging; Embassy offers no equivalent runtime memory-usage API, making static BSS analysis the primary observability mechanism. Fourth, application-layer source code size was comparable in both implementations, at approximately 1,200–1,400 source lines of code (SLOC) for the non-driver application layer, indicating that the memory savings are architectural rather than due to code brevity.

### 4.9 Limitations

Several limitations apply. First, metrics were collected on a single hardware specimen; device-to-device variation was not characterised. Second, this study focused on static (compile-time) memory and did not measure runtime stack high-water marks or interrupt stack usage. Third, execution-time performance (interrupt latency, Ethernet throughput, task jitter) was not measured. Fourth, the Embassy HAL for CH32V203 is an unofficial fork; as the upstream matures, binary sizes may change. Fifth, the relative SRAM advantage will differ on devices with larger SRAM where lwIP pool saturation is less critical. Sixth, only single compiler versions were used for each toolchain (GCC 8.2.0 for FreeRTOS-C; stable Rust, edition 2021, for Embassy-Rust); alternative versions or optimization settings could alter binary sizes by up to 5–15%. Seventh, Embassy's cooperative scheduler is sufficient for the studied IoT workload, but applications requiring hard real-time guarantees at the microsecond level may still benefit from FreeRTOS's preemptive scheduling model — a trade-off not quantified here.

## 5 Conclusion

This paper introduced a per-symbol attribution method that decomposes the static RAM of a linked firmware image into architectural categories, and applied it to sixteen builds spanning eight cumulative milestones implemented twice, in FreeRTOS-C and in Embassy-Rust, on a 64 KB RISC-V microcontroller with on-chip Ethernet. Because the categories partition the same address ranges the size utility measures, their sums reconcile exactly with the segment totals for every build, which makes the decomposition a measurement rather than an estimate. The principal findings are as follows.

First, Embassy-Rust consumed substantially less RAM than FreeRTOS-C at every milestone. At the full MQTT milestone M07, Embassy required 14,568 bytes of RAM (22.2% of capacity) versus 63,776 bytes for FreeRTOS-C (97.3%), a reduction of 49,208 bytes (77.2%). The FreeRTOS-C stack left only 1,760 bytes free, which is insufficient for TLS or additional application logic; Embassy-Rust left 50,968 bytes free.

Second, the attribution shows that both stacks commit nearly the same proportion of their RAM to networking (63.8% against 63.9% at M07) while differing 4.4-fold in absolute cost, 40,719 bytes against 9,316 bytes. The stacks also scale along different variables: FreeRTOS-C task stacks grow linearly at roughly 1 KB per task, whereas Embassy-Rust async state grows with state-machine complexity. That distinction, rather than any device-specific constant, is what should be expected to carry to other platforms.

Third, the no-heap discipline could not be sustained symmetrically. The lwIP port layer requires the FreeRTOS dynamic API, forcing a bounded arena of 14 KB at M05 and 20 KB at M06-M07 — 31.3% of the device's SRAM at M07 — while Embassy-Rust's smoltcp remained allocator-free at every milestone. This asymmetry is a property of the network stacks, not a configuration choice.

Fourth, Flash consumption converged at higher milestones: at M07, FreeRTOS-C used 76,824 bytes (58.6%) and Embassy-Rust 73,164 bytes (55.8%), a difference of only 3,660 bytes. Flash is therefore not the binding constraint; SRAM is.

Fifth, the build-time comparison depends entirely on the regime measured. In cold builds that recompile every dependency, FreeRTOS-C is faster at every milestone: Embassy-Rust takes 2.4-2.6x longer before networking is added and 10-12% longer at the networked milestones. In warm incremental builds, the regime of day-to-day development, Embassy-Rust is 27-75x faster. Both figures are medians of five sequential runs; reporting either alone would misstate the result.

These results confirm that Embassy-Rust is a practical and demonstrably superior choice for deploying full IoT networking stacks on RISC-V microcontrollers with 64 KB or less of SRAM. For the CH32V203 class of devices, FreeRTOS-C with lwIP requires careful lwIP pool minimization and leaves almost no headroom; Embassy-Rust provides comfortable memory margins even for future extensions such as TLS, additional sensors, or OTA firmware update.

Future work could extend this comparison in several directions. First, runtime execution performance — interrupt latency, Ethernet throughput, and task jitter — was not measured and should be characterised to complete the picture. Second, actual TLS integration under Embassy-Rust using embedded-tls or WolfSSL in static memory mode would validate the feasibility estimates in Table 9. Third, a migration effort study could quantify the developer-hours required to port an existing FreeRTOS-C codebase to Embassy-Rust on the same hardware. Fourth, replicating the eight-milestone methodology on the CH32V307 (128 KB SRAM, same RISC-V/Ethernet family) and the ESP32-C3 (400 KB SRAM, RISC-V, Wi-Fi) would validate generalization across the RISC-V device family. Fifth, power consumption during active networking and idle periods was not measured; Embassy's event-driven async model is hypothesised to enable earlier and longer sleep states than FreeRTOS's tick-driven scheduler. Sixth, measuring interrupt response latency using an oscilloscope would resolve the open question about the CH32V203's PFIC hardware stack versus a standard RISC-V CLINT implementation.

## Statements and Declarations

**Funding.** This research received no specific grant from any funding agency in the public, commercial, or not-for-profit sectors.

**Competing Interests.** The authors declare that they have no competing interests.

**Ethics Approval.** Not applicable. This study involved no human participants or animals.

**Consent.** Not applicable.

**Availability of Data and Materials.** All measurement data supporting the findings of this study are included with the submission: per-milestone binary metrics for both build regimes (build_metrics_coldbuild.csv, build_metrics_warmbuild.csv, holding the medians of five sequential runs, together with the ten raw run files) and the per-symbol attribution output (memory_decomposition.csv). The decomposition can be regenerated from the linked firmware images with the tool described in Section 3.8.

**Code Availability.** The sixteen firmware implementations (eight milestones in each of FreeRTOS-C and Embassy-Rust), the automated build-and-measure harness, and the per-symbol attribution tool are available from the corresponding author on reasonable request. The local ch32-hal timer patch described in Section 3.3 is included with them; an equivalent correction was independently proposed to the upstream project while this work was in preparation, so the patch is a build workaround rather than a claimed upstream contribution.

**Authors' Contributions.** R.F. designed the study, implemented both firmware stacks, developed the measurement harness and the attribution method, performed the analysis, and wrote the manuscript. J.E.R. supervised the research, advised on the experimental design, and reviewed the manuscript. W.R.P.U. and Z.M.A. contributed to the hardware setup and functional verification and reviewed the manuscript. All authors read and approved the final manuscript.

**Use of AI Tools.** During the preparation of this work the authors used an AI assistant to format and assemble the manuscript into the journal template (structure, citation and reference formatting, and layout). It was not used for research design, firmware implementation, data collection, measurement, or figure generation. The authors reviewed and edited the content and take full responsibility for the publication.